\documentstyle[emulateapj,psfig]{article}
\newcommand{\be}{\begin{equation}}
\newcommand{\ee}{\end{equation}}
\newcommand{\ba}{\begin{eqnarray}}
\newcommand{\ea}{\end{eqnarray}}
\newcommand{\siml}{\lower4pt \hbox{$\buildrel < \over \sim$}}
\newcommand{\simg}{\lower4pt \hbox{$\buildrel > \over \sim$}}

\def\xmm{{\it XMM}-Newton}

\slugcomment{Submitted to ApJ Letters}

\begin{document}

\title{XMM-Newton observation of 
the drifting pulsar B0943+10}

\author{Bing Zhang$^1$, Divas Sanwal$^2$ \& George G. Pavlov$^2$}
\affil{$^1$ Department of Physics, University of Nevada, Las Vegas,
Las Vegas, NV 89154; bzhang@physics.unlv.edu \\
$^2$ Department of Astronomy \& Astrophysics, Pennsylvania State
University, 525 Davey Lab., University Park, PA 16802;
divas@astro.psu.edu, pavlov@astro.psu.edu} 

\begin{abstract}
Radio pulsar subpulse drifting has been interpreted as rotation of
sub-beams (sparks) of pair plasma produced by intermittent breakdowns
of an inner vacuum gap above the pulsar polar cap.  This model also
predicts strong thermal X-ray emission from the polar cap caused by
inflowing particles created in spark discharges.  We have observed the
best-studied drifting pulsar B0943+10 with {\sl XMM-Newton} and
detected a point source coincident with the radio pulsar position.
Its spectrum could be fitted with a thermal blackbody model, although
a power-law model is also acceptable.  The thermal fit gives a
bolometric luminosity $L_{\rm bol}\approx 5 \times 10^{28}\,{\rm
erg\,s^{-1}}$ and a surface area $A \approx 10^3 (T/3\,{\rm
MK})^{-4}\,{\rm m}^2$, much smaller than the conventional polar cap
area, $6\times 10^4\,{\rm m}^2$.  Such thermal
radiation can be interpreted as emitted from footprints of sparks
drifting in an inner gap of a height $h \sim0.1$--$0.2\, r_{\rm pc}$, where
$r_{\rm pc}$ is the polar cap radius. However, the original vacuum
gap model by Ruderman and Sutherland requires some modification
to reconcile the X-ray and radio data.
\end{abstract}

\keywords{pulsars: individual (PSR B0943+10) - stars: neutron -
X-rays: stars}

\section{Introduction}

The mechanism of radio pulsar emission remains a mystery after decades
of study (Melrose 2004). It is generally agreed that there exists an
inner magnetospheric charge acceleration region (or inner gap) near
the pulsar polar cap.  Ruderman \& Sutherland (1975, hereafter RS75)
proposed that the gap is nearly vacuum, based on the assumption that
ions cannot be stripped off the surface by the electric field
component parallel to the magnetic field lines.  Such an assumption
was disputed by later workers (e.g., Jones 1986; K\"ossl et al.\
1988), who argued that the surface charges can be freely supplied into
the magnetosphere.
The inner accelerator flow is therefore expected to
be space charge limited (Arons \& Scharlemann 1979; Harding \&
Muslimov 1998,2001,2002; Harding et al.\ 2002). Nevertheless, the
vacuum gap model is widely used to interpret radio data. Suggestions
to amend the binding energy problem have been made (e.g., Usov \&
Melrose 1996; Xu et al.\ 1999; Gil \& Melikidze 2002), and some radio
emission models heavily rely upon the existence of such an inner
vacuum gap (e.g., Qiao \& Lin 1998; Melikidze et al.\ 2000).  To
resolve this controversy, one should look for independent
observational signatures of the vacuum gaps.

A distinguishing property of the vacuum gap model is its high polar
cap heating rate. In this model, the vacuum gap breaks down
intermittently due to pair production discharges (RS75), and the
number of the outflowing particles is comparable to that of the
inflowing particles, which results in a high luminosity of polar caps
heated up to X-ray temperatures (e.g., Zhang et al.\ 2000). On the
other hand, the space-charge-limited flow (SCLF) model predicts a much
lower polar cap heating rate because only a small fraction of the
positrons (for an electron accelerator) turn around and bombard the
surface (Arons \& Scharlemann 1979; Zhang \& Harding 2000; Harding \&
Muslimov 2001,2002). Therefore, measuring the thermal X-ray luminosity
from heated polar caps can provide a clue to verify the existence of
the vacuum gaps.

The strongest support for the vacuum gap model comes from its ability
to explain the regular sub-pulse drifting observed in some
long-period, old pulsars with the so-called ``conal'' emission beam
(Rankin 1986, and references therein). The phenomenon is naturally
interpreted as the ${\bf E \times B}$ drift of polar cap ``sparks''
circulating around the magnetic pole (RS75). One of the best-studied
drifting pulsars is PSR B0943+10 ($P=1.10$ s, $\dot{E}=1.0\times
10^{32}$ erg s$^{-1}$, $\tau = 5.0$ Myr, $B_p=4.0\times 10^{12}$ G).
Deshpande \& Rankin (1999, 2001) monitored the drifting pattern and
revealed the ``polar cap map'' of this pulsar. They identified 20
sparks rotating counterclockwise with a period of 37 rotation periods
(i.e., $\hat P_3 \simeq 37 P$, where $P$ is the spin period, and $\hat
P_3$ is the minimum period for the drifting pattern to repeat itself),
a conclusion confirmed by an analysis of low-frequency radio data
(Asgekar \& Deshpande 2001). The result was claimed to be consistent
with the RS75 model (e.g., Deshpande \& Rankin 1999, 2001; Gil \&
Sendyk 2003), although some modifications to the model may be needed
(Gil et al.\ 2002; Asseo \& Khechinashvili 2002; Gil et al.\ 2003). 

According to the Galactic electron density model (Cordes \& Lazio
2002), the distance to PSR~B0943+10 determined from its dispersion
measure (DM$ =15.35 ~{\rm pc\,cm^{-3}}$) is $0.63\pm 0.10$ kpc.  The
expected high X-ray luminosity and the small distance make it the best
target to test the predictions of the vacuum gap polar cap heating
model.  Here we report on observations of PSR B0943+10 with {\sl
XMM}-Newton, aimed at detecting the thermal X-ray emission from the
polar cap region.

\begin{figure*}[ht]
\hskip 0.3cm
\psfig{figure=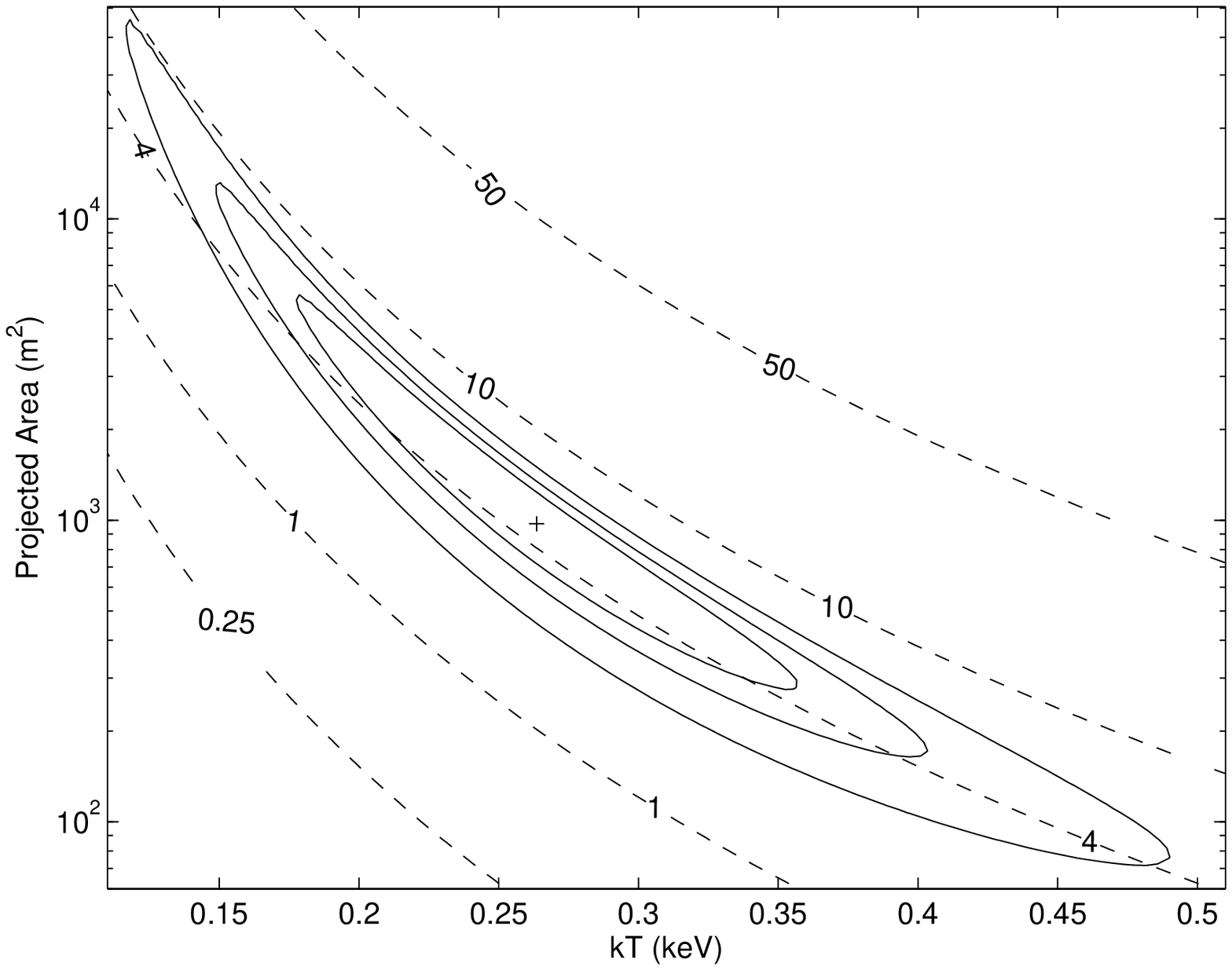,height=6.9cm,angle=-90}
\psfig{figure=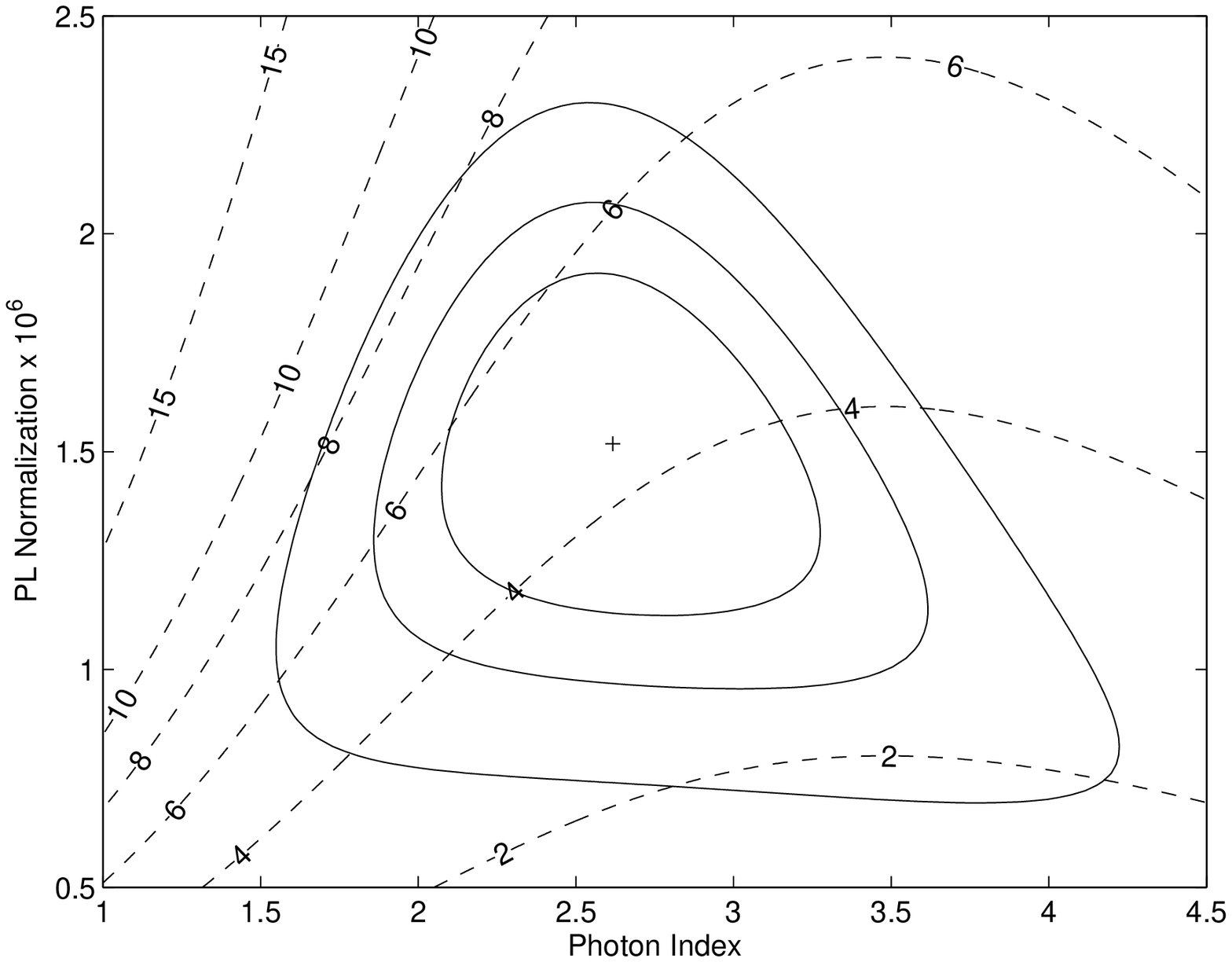,height=6.9cm}
\caption{\footnotesize{
Confidence contours (68\%, 90\%, and 99\%) for the blackbody (left)
and power-law (right) model fits to the EPIC-pn spectrum of PSR
B0943+10.  The BB normalization (vertical axis) is the projected
emitting area in units of m$^2$, for $d=630$ pc.  The PL normalization
is in units of $10^{-6}$ photons cm$^{-2}$ s$^{-1}$ keV$^{-1}$ at 1
keV.  The lines of constant bolometric luminosity (left panel; for a polar
cap close to the center of stellar disk, in units of $10^{28}$ erg
s$^{-1}$) and constant flux (right panel; in units of $10^{-15}$ erg
cm$^{-2}$ s$^{-1}$) are plotted as dashed lines, for fixed $n_{\rm H}
=4.3\times 10^{20}$ cm$^{-2}$.
}} 
\label{fig1}
\vskip -0.2cm
\end{figure*}

\section{Observations and Data Analysis }

PSR B0943+10 was observed twice with the European Photon Imaging
Camera (EPIC) aboard \xmm.  The EPIC-pn (hereafter PN) and two
EPIC-MOS (hereafter MOS) detectors were operated in extended full
frame and full frame modes, respectively.  Thin filters were used for
all the three EPIC detectors.  The initial observation of 2003 May 7
was interrupted by a strong solar flare.  The exposure time for this
observation, compromised by a high background, was 15 ks.  Second
observation was taken on 2003 December 2--3, for 20 ks.  We processed
the data using \xmm\ Science Analysis Software (SAS), v.\ 6.0.  We
removed the intervals of high background and retained only patterns
0--4 and 0--12 for the PN and MOS detectors, respectively.  We also
applied energy filters to keep the events only in the energy range
0.5--8.0 keV.  The resulting good exposure times are 24.3 ks for PN,
32.8 ks for MOS1, and 32.9 ks for MOS2.

We created a composite image using all the data (PN+MOS) in the 0.5--8
keV band.  Using a modified version of the CIAO {\em wavdetect}
tool\footnote{{\tt http://cxc.harvard.edu/ciao/threads/wavdetect}}, we
found 22 point sources in the $20'\times 20'$ field.  We searched for
counterparts of the field sources in the 2MASS and USNO-B1.0 catalogs
and found three matches in 2MASS and one match in USNO-B1.0, within
$2''$ of the X-ray source positions.  (There are no other matches
closer than $6''$.)  Assuming these are the counterparts of the X-ray
sources, we found average offsets of the X-ray with respect to
optical/NIR positions of $+0\farcs1$ in RA\ and $-0\farcs9$ in Dec,
with an rms error of about $0\farcs5$ in each coordinate (which does
not include the uncertainty of the X-ray source position
determination).  Visual inspection of the image shows a faint source
at the (boresight-corrected) position $\alpha = 09^{\rm h}46^{\rm
m}07\fs 4$, $\delta = 09^\circ 51'54''$ (J2000.0), with an uncertainty
of about $5''$ in each coordinate. The main source of the uncertainty
is a large error in centroiding the very faint X-ray source.  Since
the separation of 2\farcs7 between the X-ray source position and the
radio pulsar position ($\alpha = 09^{\rm h}46^{\rm m}07\fs 652$,
$\delta = 09^\circ 51' 55\farcs52$) is smaller than the source
position uncertainty, we conclude that the observed source is most
likely the X-ray counterpart of the radio pulsar.

We measured background-subtracted source count rates of $2.52\pm0.48$,
$0.95\pm0.19$, and $1.03\pm0.21$ counts ks$^{-1}$ for the PN, MOS1,
and MOS2 detectors, respectively.  These count rates were measured in
the 0.5--8 keV band within a $16''$ radius aperture, which contains
$\approx 67\%$ and 71\% of total point source counts for the PN and
MOS detectors, respectively.

We were unable to find a satisfactory way to combine the PN and MOS
data for spectral analysis of this faint source as the responses of
these detectors are very different.  Since the number of source counts
in the MOS data is very small, we used only the data from the most
sensitive PN detector.  We extracted the source+background spectrum
(102 counts) for the combined PN data from a $16''$ radius circle and
the background from a nearby $32''$ radius source-free region.  We
binned the spectrum to have a minimum of 10 counts per energy bin.
Spectral fitting was done using the XSPEC package (v.\ 11.3.0).

Because of the scarce count statistics, we fixed the absorbing column
to the value corresponding to the pulsar's DM (assuming 10\% ionized
ISM): $n_{\rm H} =4.3\times10^{20}$ cm$^{-2}$.  The spectrum fits
equally well with an absorbed blackbody (BB) or an absorbed power-law
(PL) model ($\chi_\nu^2 = 0.9$ and 0.7, respectively, for 8 dof).  The
BB model gives a best-fit temperature $T = 3.1$ MK and a projected
emitting area of $A\sim 10^3 d_{630}^2$ m$^2$, corresponding to an
effective radius of about $18 d_{630}$ m of an equivalent emitting
disk, where $d_{630} = d/(630\,{\rm pc})$.  
This thermal radiation could be interpreted as
emitted from a hot polar cap; in this case the bolometric luminosity
is $L_{\rm bol} = A
% \pi r_{\rm pc}^2
\sigma T^4 = 4.9^{+0.6}_{-1.6}\times
10^{28} d_{630}^2 \langle \cos \theta\rangle^{-1}$ erg s$^{-1}$, where
$\langle \cos \theta\rangle$ is a time-averaged cosine of the angle
between the magnetic axis and the line of sight ($\langle \cos
\theta\rangle = 0.97$ for the axis orientations inferred by Deshpande
\& Rankin 2001). The errors here and below are determined from the
68\% confidence contours for joint variation of two fitting
parameters.  Although the strongly correlated temperature and radius
vary in broad ranges, the luminosity is much more restricted, as
demonstrated in the left panel of Figure~1.

The PL model gives a photon index
$\Gamma = 2.6_{-0.5}^{+0.7}$ and 
a flux $F_{0.5-8\,{\rm keV}} = 
4.4^{+1.8}_{-1.5}\times 10^{-15}$ erg cm$^{-2}$
s$^{-1}$.
This flux corresponds to 
an isotropic luminosity $L_{0.5-8\,{\rm keV}}=
4\pi d^2 F_{0.5-8\,{\rm keV}}^{\rm unabs}=
2.4^{+0.8}_{-0.7}\times 10^{29} d_{630}^2$
erg s$^{-1}$. 

The inferred parameters of the X-ray source do not contradict to the
hypothesis that it is the counterpart of the pulsar (see \S3).  The
most direct way to prove the identification would be detection of
X-ray pulsations. Our timing analysis of the PN data (time resolution
0.2 s) did not reveal any significant signal at the pulsar's period of
1.1 s.  This result, however, is not restrictive: the radiation should
be about 100\% pulsed to find pulsations with only 102 counts detected
in two observations separated by about 7 month (of which only about
61 counts are expected to come from the source).  Thus,
given the uncertainty of the spectral results and a lack of period
detection, we cannot completely rule out the possibility that we
detected a field X-ray source very close to the pulsar position.  In
this case, the above flux/luminosity estimates should be considered as
upper limits.

\section{Physical implications}

As described in \S2, the quality of the data is too poor to
discriminate between the thermal (BB from a polar cap) and nonthermal
(PL from the magnetosphere) spectral fits of the putative X-ray
counterpart of PSR B0943+10.  The photon index for the PL fit is
somewhat larger than, but not inconsistent with, those observed in
other pulsars ($\Gamma=1$--2), given its large uncertainty (in
particular, $\Gamma$ decreases if a smaller $n_{\rm H}$ value is
assumed). The luminosity found from the PL fit allows one to estimate
the ``X-ray efficiency'' of the magnetospheric emission, $\eta \equiv
L_X/\dot{E}$. It follows from the above results that $\eta =2.4 
(-1.2,+1.8)\times 10^{-3}$ and $0.8 (-0.6,+1.5)\times 10^{-3}$ in the
0.5--8 and 2--10 keV bands, respectively, where the distance
uncertainty is included (the latter band is often used for
characterization of the nonthermal pulsar radiation; e.g., Possenti et
al.\ 2002).  Although the inferred efficiency
is rather high (close to the upper limit suggested by Possenti et al.\
2002), it is within the range of efficiencies found recently for other
old pulsars (Zavlin \& Pavlov 2004).

Within the thermal interpretation of the observed radiation,
particularly interesting implications follow from the comparison with
the predictions of polar cap heating models, especially the vacuum gap
model (RS75). In this model, a vacuum gap of height $h$ breaks down
due to run-away electron-positron pair production, forming ``sparks''
along the magnetic field lines.
A half of the relativistic particles produced in the spark discharge
stream back toward the neutron star and heat the ``spark footprints''
on the polar cap surface.  Since the strong magnetic field suppresses
the heat conduction across the field lines, and the cooling timescales
are very short (about a few $\mu$s for $B=$ a few $\times
10^{12}$\,G; Gil et al. 2003), only the spark footprints emit thermal
X-rays while the rest of the polar cap remains cold.  The BB fit of
the observed spectrum allows one to estimate the bolometric
luminosity, $L_{\rm bol} \approx 5\times 10^{28}$ ergs s$^{-1}$, and
the total X-ray emitting area that is strongly correlated with the
temperature, $A \approx 10^3 (T/3\,{\rm MK})^{-4}$ m$^2$ ($A \approx
300$--5000 m$^2$, $T\approx 2.0$--4.2 MK, at 68\% confidence).  Even
the largest area allowed by the fit is much smaller than the
conventional polar cap area for a dipole magnetic field, $A_{\rm
pc}^{\rm dip} =\pi (r_{\rm pc}^{\rm dip})^2 \simeq 6\times 10^4$
m$^2$, where $r_{\rm pc}^{\rm dip} = (2\pi R^3/c P)^{1/2} \simeq 138$
m, for a neutron star radius $R=10$ km.  The spark footprints fill a
fraction of the polar cap, with a filling factor $f=A/A_{\rm pc} \sim
0.017 b A_3 $, where $A_3 = A/(10^3\,{\rm m}^2$), and 
the factor $b = B_s / B_p \sim
A_{\rm pc}^{\rm dip}/A_{\rm pc} >1$
takes into account that the magnetic field is non-dipolar
(i.e., the field in the gap, $B_s$, may exceed the dipolar value $B_p$,
and the polar cap area may be smaller
than $A_{\rm pc}^{\rm dip}$; Gil et al.\ 2002, 2003).
The radio sub-pulse drifting was interpreted as caused by 20
sparks drifting along the circular outer boundary of the polar cap
(Deshpande \& Rankin 1999,2001). Since some sparks may not be seen in
the radio (but should be seen in more isotropic thermal X-rays), the
total number of sparks $N$ can exceed 20 (e.g., Gil \& Sendyk 2000
suggest $N\simeq 42$).
A characteristic spark radius, $r_{\rm sp}$, can be estimated as
(assuming all sparks have the same size)
\be
r_{\rm sp} \sim (A/\pi N)^{1/2} = 4\, A_3^{1/2} (N/20)^{-1/2} {\rm m}.
\label{rsp}
\ee

Another constraint on the gap and spark parameters
comes from the comparison of the bolometric
and spin-down luminosities. The bolometric 
polar cap luminosity can be 
estimated as $L_{\rm bol} \sim
(1/2) \Phi c
\rho_{_{\rm GJ}} A$,
where $\Phi \sim 2\pi B_s h^2/cP$ is the
potential drop across the gap (along the magnetic field lines), and
$\rho_{_{\rm GJ}} 
%\propto B_s$ 
\sim B_s/cP$
is the Goldreich-Julian charge density.
On the other hand, the total spin-down luminosity 
can be written as $\dot E \sim 2 \Phi_{\rm max} c \rho_{_{\rm
GJ}}^{\rm dip} A_{\rm pc}^{\rm dip}= 2 \Phi_{\rm max} c \rho_{_{\rm
GJ}} A_{\rm pc}$, 
% \pi
%r_p^2$, 
where 
$\Phi_{\rm max} = B_p A_{\rm pc}^{\rm dip}/cP = B_s A_{\rm pc}/cP$
is the maximum possible
potential drop along a magnetic field line (RS75). This gives 
$L_{\rm bol}/\dot{E} \sim (f/4) (\Phi/\Phi_{\rm max}) =
(1/2) N (r_{\rm sp}/r_{\rm pc})^2 (h/r_{\rm pc})^2$. 
Using the observational value $L_{\rm bol}/\dot{E}\approx 5\times 10^{-4}$,
we obtain 
\be
(h/r_{\rm pc}) (r_{\rm sp}/r_{\rm pc}) \sim 7\times 10^{-3}
(N/20)^{-1/2}. 
\label{hrsp/rpc2}
\ee
Adopting the assumption that the size of the spark is
comparable to the gap height 
(i.e., $r_{\rm sp} \sim h/2$; Gil \& Sendyk 2000), 
we obtain from equation (2)
\be
h/r_{\rm pc} \sim 0.12\, (N/20)^{-1/4}\,.
\label{h/rpc}
\ee
This gives $h \sim 16\, b^{-1/2} (N/20)^{-1/4}$ m, 
and $r_{\rm sp} \sim 8\, b^{-1/2} (N/20)^{-1/4}$ m.
Using this estimate for $r_{\rm sp}$ and equation (1), we obtain
$A \sim 4000 b^{-1} (N/20)^{1/2}$ m$^2$, which is consistent with the
X-ray data at reasonable values of $b$ and $N$ (see Fig.\ 1).

Further constraints can be obtained from the drifting period.
According to RS75, for sparks at the radial distance $r_\perp$
from the pole,
it can be estimated as $\hat{P}_3 \simeq 2\pi r_\perp/v
\simeq 2\pi r_\perp B_s/E_\perp c$, where 
$v=|{\bf E \times B_s}| c/B_s^2$ is the drift velocity,
and $E_\perp = |dV(r_{\perp})/dr_\perp|$ is the electric field component
perpendicular to the magnetic field lines at the top of the gap.
In the case $h\ll r_{\rm pc}$, $V(r_\perp)$ is approximately uniform
throughout the polar cap except for near the cap edge, $r_{\rm
pc}-r_\perp \lesssim h$ (Appendix II of RS75). This means that the
``inner sparks'' drift very slowly,
while  $E_\perp \sim \Phi/h$, $v\sim 2\pi h/P$, and
$\hat{P}_3/P \sim r_{\rm pc}/h$ for the sparks near the cap edge.
The measured $\hat{P}_3/P \simeq 37$ thus gives $h/r_{\rm pc} \sim
0.027$, or $h\sim 4\, b^{-1/2}$ m
and $r_{\rm sp} = 36\, b^{-1/2} (N/20)^{-1/2}$ m (from eq.\ [2]).
Such a spark radius substantially exceeds the gap height, and it
corresponds to the emitting area, $A \sim 8\times 10^4 b^{-1}$ m$^2$,
well above the range allowed by the X-ray fit. 
Therefore, the original vacuum gap model of RS75 can hardly be
reconciled 
with the X-ray data. However, we cannot rule out that a modification of the
RS75 model could explain both the X-ray and radio data in a consistent way.
For instance, Gil et al.\ (2003) suggested a modification assuming
thermal emission of charged particles from the polar cap surface
which can partially screen the gap potential: $\Phi \to \zeta\Phi$, where
$\zeta < 1$ is the screening parameter. In such a modified RS75 model,
$\hat{P}_3/P \sim r_{\rm pc}/\zeta h$.
For $\zeta = 0.17$, inferred by Gil et al.\
for PSR B0943+10, we obtain $h/r_{\rm pc} \sim 0.16$, very
close the value inferred from the X-ray data (eq.\ [\ref{h/rpc}]).
However, a more detailed study of the dynamics of a partially screened
unsteady polar gap is desirable for 
an accurate treatment of the problem.
We should also caution that the BB model used for
fitting the spectrum gives only crude estimates for the area and
temperature in the case of strong magnetic
fields (Zavlin \& Pavlov 2004). 
To conclude, the thermal interpretation of the X-ray data
is not inconsistent with the general idea that the X-ray radiation
is emitted from spark footprints, but the original vacuum gap model
of RS75 needs some modification to explain both the X-ray and radio
data.

The results of our X-ray observations also constrain other models.
Motivated by solving the binding energy problem, Xu et al.\ (1999)
suggested that PSR 0943+10 and other drifting pulsars might be strange
quark stars covered by a very thin layer of normal matter.  In this
case, the heat deposited at the polar cap spreads over the entire star
surface because of a high thermal conductivity of the surface layer
(Xu et al.\ 2001). This results in a nearly uniform, low surface
temperature ($\lesssim 0.1$ MK), being inconsistent with the thermal
interpretation of the observed X-ray radiation (but of course this
hypothesis cannot be excluded if the observed radiation is
nonthermal).

Polar cap heating in the SCLF model has been studied by Harding \&
Muslimov (2001,2002). Assuming a dipole field configuration near the
surface, the timing parameters of PSR B0943+10 indicate that the gap
is controlled by inverse Compton scattering (ICS) rather than
curvature radiation (CR), and that the gap is in the ``saturated''
regime. Using eqs.\ (64) and (66) of Harding \& Muslimov (2002), the
predicted polar cap luminosity for this pulsar is $8.8(1.7) \times
10^{27}\,{\rm erg\,s^{-1}}$ for a resonant (non-resonant) ICS gap.
Since these luminosities are much lower than observed, the SCLF model
is inconsistent with the thermal interpretation of the X-ray data,
but is consistent with the nonthermal interpretation.  If future
observations show that the X-ray radiation is, in fact, predominantly
nonthermal ($L_{\rm bol}\ll 5\times 10^{28}$ ergs s$^{-1}$), then the
regular drifting behavior observed in PSR B0943+10 and other drifting
pulsars will have to be interpreted by models other than bunched
plasma sparks in the inner gap (e.g., Kazbegi et al.\ 1996; Wright
2003).

\section{Summary}
To test the existence of an inner vacuum gap in PSR B0943+10, we
observed the pulsar with {\sl XMM-Newton} and detected a point source
at the radio pulsar position.  The best-fit 0.5--8 keV band isotropic
luminosity is $\sim 2.4\times 10^{29}~{\rm erg~s^{-1}}$ for the PL
fit, while the bolometric BB luminosity is about $5\times 10^{28}$ erg
s$^{-1}$.  Within the thermal interpretation, the X-ray radiation is
emitted from a heated area much smaller than the conventional polar
cap area. In the framework of the sparking pulsar model, this implies
that only spark footprints are heated up to X-ray temperatures while
the rest of the polar cap remains relatively cold.  The thermal model,
supplemented with the results of observations of subpulse drifting, is
consistent with the presence of an inner gap of a height $h \sim
0.1$--$0.2\, r_{\rm pc}$, and $N\geq 20$ rotating sparks whose
footprints fill $\sim 0.04$--0.07 of the polar cap area.  The thermal
model disfavors the strange quark star model of PSR 0943+10, and it is
also inconsistent with the SCLF polar cap heating model. Both models
are, however, allowed if the spectrum is non-thermal.

The nonthermal (magnetospheric) interpretation of the observed radiation
cannot be ruled out. In this interpretation, the X-ray efficiency,
$\eta\sim 10^{-3}$ in the 2--10 keV band, is comparable to that of
other old pulsars. Much deeper observations are required to
firmly establish the nature of the X-ray radiation from PSR B0943+10. 

\acknowledgements
This work was supported by NASA grants NAG5-13539 and NAG5-10865.


\begin{thebibliography}{}
\bibitem[]{} Arons, J. \& Scharlemann, E. T. 1979, ApJ, 231, 845
\bibitem[]{} Asgekar, A. \& Deshpande, A. A. 2001, MNRAS, 326, 1249
\bibitem[]{} Asseo, E. \& Khechinashvili, D. 2002, MNRAS, 334, 743
\bibitem[]{} Cordes, J. M. \& Lazio, T. J. W. 2002, astro-ph/0207156
\bibitem[]{} Deshpande, A. A. \& Rankin, J. M. 1999, ApJ, 524, 1008
\bibitem[]{} -----. 2001, MNRAS, 322, 438
\bibitem[]{} Gil, J. \& Melikidze, G. I. 2002, ApJ, 577, 909
\bibitem[]{} Gil, J., Melikidze, G. I. \& Geppert, U. 2003, A\&A, 407, 
315
\bibitem[]{} Gil, J., Melikidze, G. I. \& Mitra, D. 2002, A\&A, 388,
246 
\bibitem[]{} Gil, J. \& Sendyk, M. 2000, ApJ, 541, 351
\bibitem[]{} -----. 2003, ApJ, 585, 453
\bibitem[]{} Harding, A. K. \& Muslimov, A. G. 1998, ApJ, 508, 328
\bibitem[]{} -----. 2001, ApJ, 556, 987
\bibitem[]{} -----. 2002, ApJ, 568, 862
\bibitem[]{} Harding, A. K., Muslimov, A. G. \& Zhang, B. 2002, ApJ,
576, 366
\bibitem[]{} Jones, P. B. 1986, MNRAS, 218, 477
\bibitem[]{} Kazbegi, A., Machabeli, G., Melikidze, G. \& Shukre, C.
1996, A\&A, 309, 515
\bibitem[]{} K\"ossl, D., Wolff, R. G., M\"uller, E. \& Hillebrandt,
W. 1988, A\&A, 205, 347
\bibitem[]{} Melikidze, G. I., Gil, J. \& Pataraya, A. D. 2000, ApJ, 
544, 1081
\bibitem[]{} Melrose, D. 2004, in Young Neutron Stars and
Their Environments (IAU Symp.\ 218, ASP Conf.\ Proc.),
eds.\ F.\ Camilo and B.\ M.\ Gaensler, p.349
\bibitem[]{} Possenti, A., Cerutti, R., Colpi, M. \& Mereghetti, S.
2002, A\&A, 387, 993
\bibitem[]{} Qiao, G. J. \& Lin, W. P. 1998, A\&A, 333, 172
\bibitem[]{} Rankin, J. M. 1986, ApJ, 301, 901
\bibitem[]{} Ruderman, M. \& Sutherland, P. G. 1975, ApJ, 196, 51 (RS75)
\bibitem[]{} Usov, V. V. \& Melrose, D. B. 1996, ApJ, 464, 306
\bibitem[]{} Wright, G. A. E. 2003, MNRAS, 344, 1041
\bibitem[]{} Xu, R. X., Qiao, G. J. \& Zhang, B. 1999, ApJ, 522, L109
\bibitem[]{} Xu, R. X., Zhang, B. \& Qiao, G. J. 2001, APh, 15, 101
\bibitem[]{} Zavlin, V.\ E., \& Pavlov, G.\ G. 2004, ApJ, 616, 452
\bibitem[]{} Zhang, B. \& Harding, A. K. 2000, ApJ, 532, 1150
\bibitem[]{} Zhang, B., Harding, A. K. \& Muslimov, A. G. 2000, ApJ,
531, L135
\end{thebibliography}
\end{document}